\documentclass[aip,jcp,amsmath,amssymb,reprint]{revtex4-1}

\usepackage{graphicx}
\usepackage{dcolumn}
\usepackage{bm}
\usepackage{braket}
\usepackage{mathtools}

\begin{document}

\title{Exciton Dissociation and Charge Separation at Donor-Acceptor Interfaces from Quantum-Classical Dynamics Simulations}

\author{Aaron Kelly}
\email{aaron.kelly@dal.ca}
\affiliation{Department of Chemistry, Dalhousie University, Halifax, Nova Scotia, Canada}

\date{\today}

 	\begin{abstract}
 	    In organic photovoltaic (OPV) systems, exciton dissociation and ultrafast charge separation at donor-acceptor heterojunctions both play a key role in controlling the efficiency of the conversion of excitation energy into free charge carriers. In this work, nonadiabatic dynamics simulations based on the quantum-classical Liouville equation, are employed to study the real-time dynamics of exciton dissociation and charge separation at a model donor-acceptor interface. Benchmark comparisons for a variety of low dimensional donor-acceptor chain models are performed to assess the accuracy of the quantum classical dynamics technique referred to as the forward-backward trajectory solution (FBTS). Although not always quantitative, the FBTS approach offers a reasonable balance between accuracy and computational cost. The short-time dynamics of exciton dissociation in related higher-dimensional lattice models for the interface are also investigated to assess the effect of the dimensionality on the first steps in the mechanism of charge carrier generation.
 	\end{abstract}

\maketitle


\section{Introduction}

The challenge of achieving optimized energy conversion efficiency in light-harvesting devices continues to inspire efforts toward a full elucidation of the detailed mechanisms of excitation energy transport and charge transfer in a broad range of systems, from organic photovoltaic solar cells \cite{tang,heeger1992} to biological photosynthetic systems\cite{engel2007,scholes2017,bredas2017}. Due to the central importance of this problem in energy capture and storage, a large number of experimental and theoretical efforts have pursued investigations in this area in recent years, and remarkable progress has been made thus far\cite{bredas2004,heeger2009,mcgehee2013,ratner2014,ostroverkhova2016}. 

In particular, organic photovoltaic (OPV) materials present an interesting and potentially useful class of light harvesting systems. Typical OPV materials have a low dielectric constant, and the binding energy of an electron-hole pair is approximately $0.2-0.5$ $eV$ \cite{drori2008,xyz2009,hallermann2010}, while the donor-acceptor pair distance is on the order of $1 nm$\cite{gregg2003}. This Coulomb binding energy is well above the thermal energy at room temperature. Hence, relaxation of this charge-transfer state, also referred to as polaron formation, competes with the processes that lead to spatial separation of the electron and hole as free carriers, and reduces the overall charge mobility\cite{chin2014}. Despite this competition, and losses on longer timescales due to charge recombination processes\cite{deibel2009,nguyen2013,friend2014}, the efficiency of charge carrier generation in OPV materials can, in fact, be remarkably high\cite{heeger2009,durrant2010,ostroverkhova2016,heeger2012,mcgehee2013,salleo2014,bakulin2019}.   

Generally, it is thought that the charge separation (CS) process in OPVs is initiated by the dissociation of a Frenkel exciton state at the donor-acceptor interface\cite{troisi2013,tamura2013jacs,bittner2014}. The exciton (XT) dissociation proceeds via an interfacial charge transfer (CT) state, in which the electron and hole are bound to the donor and acceptor at the interface, before spatially separated charge transfer states, and eventually free carriers, are formed. Spectroscopic studies have established that the CS process can proceed on an ultrafast timescale, ranging from a few tens to a few hundreds of femtoseconds \cite{ohkita2008,xyz2008,xyz2009,grancini2013,hayes2014}. A number of plausible mechanisms have been proposed to explain how the Coulomb barrier can be surmounted on such a short timescale. One prominent proposal, referred to as hot exciton dissociation\cite{xyz2008,grancini2013,rossky2013}, is based on the excess energy of the Frenkel exciton, while another proposal is based on quantum delocalization of the charge transfer states\cite{friend2012,nelson2015,ishizaki2009,kato2018}. In addition, hybrid mechanisms that combine these themes have also been proposed\cite{tamura2013jacs}. However, the mechanisms of XT dissociation and charge separation process at OPV heterojunctions are still a matter of some debate, and the details of how this process proceeds remains the subject of active investigation.

Further developing a theoretical understanding of the XT dissociation and CS process requires both thermodynamic calculations of the free energy landscape of the system, and real-time quantum dynamics simulations. Thus far, from the thermodynamic perspective, studies have established that charge separation is entropically favourable due to the increased density of CS states\cite{gregg2003,gregg2011,hood2016}.  Efforts to simulate the real-time quantum dynamics of this problem span over more than a decade\cite{troisi2006,tamura08,troisi2011,tamura2012,chin2014,chin2015}. Using two and three state models, it was observed that XT dissociation to the CT state involves coherent electronic evolution, and hence the early steps of the process do not obey a classical incoherent rate law\cite{tamura08,tamura2012}. More recently, multiple-site lattice models\cite{jang2008,troisi2013,tamura2013jacs} have been employed for studies of model OPVs using both highly accurate quantum dynamics approaches such as the multiconfigurational time-dependent Hartee (MCTDH) method\cite{tamura2013jacs,burghardt2015,burghardt2018}, the hierarchical equations of motion (HEOM) approach\cite{tanimura1989, tanimura2006,ishizaki2005,yan2005,shi2009,kato2018,shi2108}, the multiconfigurational Ehrenfest method\cite{ma2018}, and the time-dependent density matrix renormalization group approach\cite{ma2016}, in addition to approximate dynamics methods such as Ehrenfest mean field theory\cite{troisi2015} and Redfield theory\cite{troisi2015}.   

In this work, we investigate the mechanism of exciton dissociation and charge separation in model OPV interfaces using a nonadiabatic dynamics simulation technique called the forward-backward trajectory solution (FBTS) of the quantum-classical Liouville equation\cite{qcle}. The FBTS algorithm can be rigorously derived from exact quantum dynamics\cite{fbts1}, is systematically improvable\cite{fbts2}, and is typically more accurate than Ehrenfest mean field theory and perturbative methods\cite{fbts2,mfgqme,pfalzgraff2019}, while retaining a reasonable computational cost. We report benchmark comparisons of FBTS simulation results with recently available, highly accurate, hierarchical equations-of-motion (HEOM) data\cite{kato2018,shi2108}, for a number of different parameter regimes in low-dimensional lattice models for the donor-acceptor interface. Further, we study related higher-dimensional interface models, that are currently less accessible to benchmark-level approaches, in order to assess the effect of dimensionality on the underlying mechanism.

The remainder of this work is organized as follows. In Sec. II the donor-acceptor interface models used in this study are described, as is the quantum-classical FBTS dynamics simulation methodology. In Sec. III the FBTS simulation results are reported and discussed, and in Sec. IV some conclusions and outlooks are presented.

\section{Theory}
In this section we describe a class of lattice models used to represent the donor-acceptor interface in OPV materials, and the quantum-classical FBTS dynamics simulation methodology, used in this study.

\begin{figure}
  \includegraphics[width=\columnwidth]{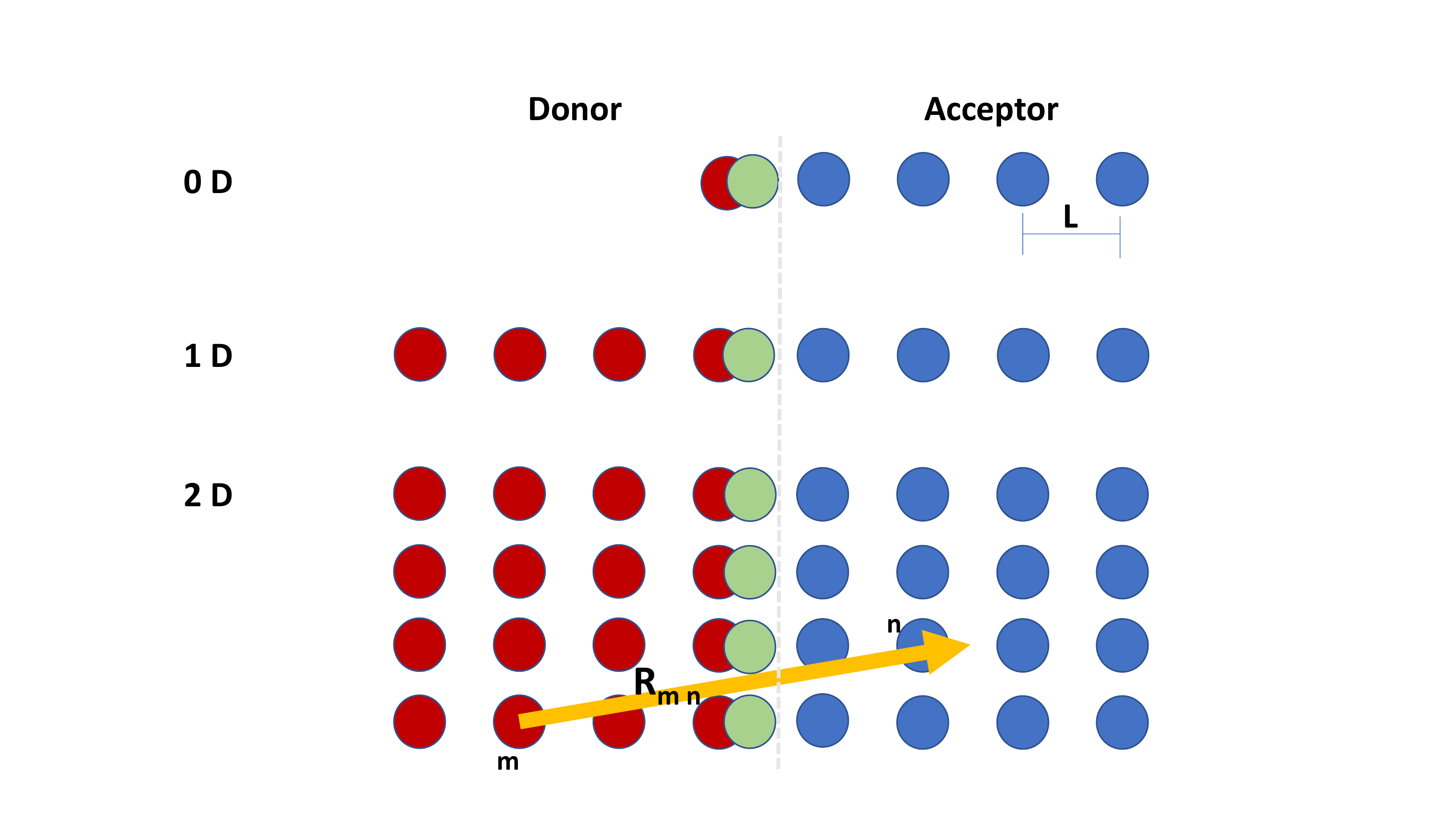}
  \caption{Cartoon representation of the lattice models for the donor-acceptor interface. The red and blue circles represent donor and acceptor molecules, and the green circles represent the excitonic states of the interfacial donor molecules. Top: zero dimensional model (0D) with single exciton and donor sites, and a chain of acceptor sites. Middle: one dimensional (1D) linear model, with multiple acceptor sites. Bottom: Two dimensional model (2D) with multiple excitonic states at the interface. The intersite vector $\mathbf{R_{mn}} $ is depicted, which measures the distance between donor site $\mathbf{m}$ and acceptor site $\mathbf{n}$. The lattice spacing in all cases is $L$.}
\label{1}
\end{figure}

\subsection{Models for the Donor-Acceptor Interface}
A class of tight-binding lattice models is used to to study XT dissociation and charge separation process, and is depicted in Fig. 1.
These models have been used to investigate both the free energy landscapes and real-time 
quantum dynamics of these processes in a number of previous studies of the donor/acceptor interface in OPV systems\cite{jang2008,troisi2013,tamura2013jacs,troisi2015,shi2108}. Each site on the lattice
represents a donor or acceptor molecule (or moiety), that has a local (on-site) set of vibrational modes that are coupled to it. 
Electronic couplings are only included for nearest-neighbour lattice sites, although in principle a more realistic 
(e.g. dipolar) form could be adopted. In Fig. 1 a simple cartoon depicts a selection of these lattice models, in increasing dimensionality. In the 'zero-dimensional' case, there is a single donor site at the interface, while in the one and two
dimensional cases there are multiple donor and acceptor sites. 

The total Hamiltonian can be written in the following general form,
\begin{equation}
    \hat{H} = \hat{H}_{el} + \hat{H}_{el-vib} + \hat{H}_{vib} +\hat{H}_{E}. 
\end{equation}
The bare electronic Hamiltonian, $\hat{H}_{el}$, describes the energies and couplings the XT states and each CT state. An XT or a CT state is denoted by the ket $|m,n\rangle$ corresponding to a pair of donor and acceptor sites, $(m,n)$, on the lattice. The electronic coupling elements are hopping-type terms that allow for the initial exciton dissociation to the interfacial charge-transfer state, as well as subsequent electron/hole transport from site to site on the lattice leading to charge separation.
\begin{eqnarray}\nonumber
    \hat{H}_{el} &=& \sum_{m,n} \epsilon_{m,n} |m,n\rangle\langle m,n| \\ &-& \sum_{m,n,k,l} \Delta_{mn,kl} (|m,n\rangle\langle k,l|+|k,l\rangle\langle m,n|), 
\end{eqnarray}
where 
\begin{equation}
    \epsilon_{m,n} = \epsilon_{m,n}^0 - \frac{e^2}{4\pi\varepsilon_{eff} R_{mn}} 
\end{equation}
where $R_{mn}$ is the distance between the electron and hole sites. The effective dielectric constant 
$\varepsilon_{eff}=\varepsilon_r\varepsilon_0$ is a product of the relative dielectric constant and the 
dielectric constant of the vacuum. For simplicity here, we assume that the donor and acceptor materials 
have the same effective dielectric constant. The coupling between charge transfer states $(m,n)$ and $(k,l)$, is $\Delta_{mn,kl}$. In the 1D model, for example, the nearest-neighbour coupling elements 
can be written in the following form,
\begin{eqnarray}
   \Delta_{mn,kl} = \Delta(\delta_{m,k\pm1}\delta_{n,l}+\delta_{m,k}\delta_{n,l\pm1}).
\end{eqnarray}

Using mass-weighted coordinates and momenta, the on-site vibrational Hamiltonian can be expressed as follows,
\begin{equation}
    \hat{H}_{vib} = \sum_m\sum_j^{N_b^m}\Big(\frac{\hat{P}_{mj}^2}{2} + \frac{1}{2}\omega_{mj}^2\hat{Q}_{mj}^2\Big),
\end{equation} where $\omega_j$ are the frequencies of each mode, $N_b^m$ is the number of modes at site $m$, and the index $m$ runs over all the sites on the lattice. 

The electron-vibrational coupling is linear in the coordinates of each vibrational degree of freedom, and only affects the relative energies of the XT and CT states (and not the couplings),
\begin{equation}
    \hat{H}_{el-vib} = \sum_{m,n}\Big(\sum_j^{N_b^m}c^{el}_{mj}\hat{Q}_{mj}+\sum_j^{N_b^n}c^{h}_{nj}\hat{Q}_{nj}\Big)| m, n  \rangle \langle m, n |.
\end{equation} The coupling constants, $c^{el}_{j}$ and $c^{h}_{j}$, set the strength of the electron-vibrational interaction and are related to the spectral density for the vibrational DOFs. The spectral density at each site is chosen to be equivalent, and of the Drude-Lorentz (Debye) form \cite{kato2018,shi2108},
\begin{equation}
    J(\omega) = \frac{2 \lambda \omega_c \omega}{\omega_c^2 + \omega^2},
\end{equation}
where $2\lambda$ is the vibrational reorganization energy, and $\omega_c$ is the inverse of the polaron formation time\cite{chin2014}. 

The final term in the total Hamiltonian, $\hat{H}_{F}$, describes coupling of the CT states to an externally applied static electric field,
\begin{equation}
    \hat{H}_{F} = -\mathit{e} \mathbf{E} \cdot \sum_{m,n} \mathbf{R}_{m, n}|m, n \rangle \langle m, n |,
\end{equation}
where $\mathbf{R}_{m, n}$ is a vector connecting donor site $m$ to acceptor site $n$ on the lattice. The electric field vector $\mathbf{E} = E \underline{\mathbf{e}}_F$, is a product of the scalar field strength $E$, and a unit polarization vector $\underline{\mathbf{e}}_F$ . In all cases where the field strength is nonzero here, it is oriented perpendicular to the donor-acceptor interface (see Fig. 1), with $E$ $=$ $10 V/\mu m$\cite{janssen08,shi2108}. 

\subsection{Quantum-Classical Liouville Theory}

In this work, an approximate quantum-classical dynamics technique that stems 
from the quantum-classical Liouville equation (QCLE)\cite{qcle,kapral2015}, that is referred 
to as the forward-backward trajectory solution (FBTS)\cite{fbts1}, is employed. A brief overview of this method is 
given here, and the interested reader is directed to the original work of Hsieh and Kapral \cite{fbts1,fbts2} for 
more detailed derivations of the expressions.

In quantum-classical Liouville theory different representations are used for the 
degrees of freedom (DOF) that are to be treated quantum mechanically, and the DOF
that are to be treated classically. While the quantum DOF are initially treated 
as abstract quantum operators, the classical degrees of freedom are cast onto a 
continuous classical phase-space description via the Wigner transform\cite{qcle}. In the class 
of problems described in the previous section, the vibrational degrees of freedom 
correspond to the classical-like system, while the exciton, electron, and hole states 
are the quantum subsystem.

In this mixed representation the total Hamiltonian is,
\begin{eqnarray}
    \hat{H}_W (X) &=& \hat{H}_{el} + \hat{H}_{el-vib,W} (R) + H_{vib,W} (R,P) \\\nonumber &=& H_{vib,W} (X) + \hat{h}(R),
\end{eqnarray}
where the subscript $W$ indicates that the partial Wigner transform over the vibrational degrees of freedom 
has been performed. The partial Wigner transform of a general operator, $\hat{B}_W(X)$, thus has both a quantum 
mechanical operator character in the electronic Hilbert space, and is a function of the phase space variables 
corresponding to the vibrational degrees of freedom, 
$X =(R,P) = (R_1, R_2,...,R_{N_b} , P_1, P_2,...,P_{N_b} )$.

The QCL evolution equation for an arbitrary quantum mechanical operator, $\hat{B}$, can be written as 
\begin{equation}
\frac{\partial \hat{B}_{W}(t)}{\partial t} = \frac{i}{\hbar}\Big[ \hat{H}_{W} , \hat{B}_{W} (t) \Big] - \frac{1}{2} \Big( \Big\{ \hat{H}_{W}, \hat{B}_{W}(t) \Big\} - \Big \{\hat{B}_{W}(t), \hat{H}_{W} \Big\} \Big),
\end{equation}
where the square and curly brackets denote the commutator and Poisson bracket, respectively. 
This expression may also be written in a more compact form, defining the quantum-classical Liouville operator $\mathcal{L}$, 
\begin{equation}
    \frac{\partial \hat{B}_W(t)}{\partial t} = i\mathcal{L}\hat{B}_W(t).
\end{equation}

The QCLE has a number of attractive features; for example, it conserves energy and phase space volumes, and is 
formally exact for the class of models for the donor acceptor interface considered in this work (an arbitrary electronic 
system that is bilinearly coupled to a harmonic environment). However, the QCLE lacks a Lie algebraic structure\cite{nielsen2001}, and 
thus systematic difficulties can be encountered in the case of more general Hamiltonians, or in constructing approximate 
solutions. Both of these aspects are the subject of current development. A careful ordering of the forward and backward time evolution operators can be imposed on the route to deriving approximate equations of motion from the QCLE, in order to help mitigate potential complications that might arise, due to the lack of Lie structure, in the resulting approximations. We outline one such approach in the next section.

\subsection{Forward - Backward Trajectory Solution}
To set the stage for the forward-backward formulation of the dynamics, note that the action of the quantum-classical 
Liouvillian, $\mathcal{L}$, on an operator $\hat{B}_W$ in equation (11) can be written in the following form,
\begin{equation}
    i\mathcal{L}\hat{B}_W = \frac{i}{\hbar}\Big(\overrightarrow{H_{\Lambda}} \hat{B}_W - \hat{B}_W \overleftarrow{H}_{\Lambda} \Big).
\end{equation}
This expression introduces forward- and backward-acting mixed quantum-classical Hamiltonian operators, 
$\overrightarrow{H_{\Lambda}} = \Big( 1+\frac{\hbar \Lambda}{2i}\Big)\hat{H}_W$, and $\overleftarrow{H_{\Lambda}} = \hat{H}_W \Big(1+\frac{\hbar\Lambda}{2i}\Big)$, and $\Lambda$ is defined to be the negative of the Poisson-bracket operator. 
Hence, the formal solution of the QCLE can be written in terms of the forward and backward acting propagators, 
\begin{equation}
\hat{B}_W(X,t) = e^{i\mathcal{L}t}\hat{B}_W(X,0) = \mathcal{S}\Big(e^{i\overrightarrow{H}_{\Lambda}t}\hat{B}_W(X,0)e^{i\overleftarrow{H}_{\Lambda}t}\Big)
\end{equation} where operator $\mathcal{S}$ ensures that the proper ordering of the forward and backward evolution operators acting on $B_W(X)$ is maintained. 

The next step is to introduce a basis representation for the subsystem degrees of freedom. In this case, 
the Meyer-Miller-Stock-Thoss (MMST) mapping representation\cite{meyermiller,stockthoss} for the subsystem basis $\{|\lambda\rangle\}$ is chosen. In the MMST representation, a subsystem state $|\lambda\rangle$ is represented by a mapping state 
$|m_{\lambda}\rangle$, that is an eigenfunction of a system N fictitious harmonic oscillators, that have 
occupation numbers which are constrained to be 0 or 1: 
$|\lambda\rangle\} \rightarrow |m_{\lambda}\rangle$ = $|0_1, ... , 1_{\lambda},... 0_N\rangle$. 

The mapping version of an operator, $\hat{B}_{m}(X)$, is defined such that its matrix elements are equivalent to those of the corresponding  operator, $\hat{B}_{W}(X)$. For example, the mapping Hamiltonian can be written as\cite{Kim2008}
\begin{equation}
\hat{H}_m(X) = H_{vib}(X) + \sum_{\lambda \lambda'} h_{\lambda \lambda'}(R) \hat{a}ˆ{\dag}_{\lambda}\hat{a}_{\lambda'},    
\end{equation}
where the creation and annihilation operators on the subsystem mapping states, $\hat{a}ˆ{\dag}_{\lambda}$ and $\hat{a}_{\lambda}$,  satisfy the usual bosonic commutation relation $[\hat{a}ˆ{\dag}_{\lambda}, \hat{a}_{\lambda'}] = \delta_{\lambda \lambda'}$.

The formal solution of the QCLE, Equation (13), can written in the mapping representation as follows,
\begin{equation}
    \hat{B}_m(X,t) = \mathcal{S}\Big(e^{i\overrightarrow{H}^m_{\Lambda}t}\hat{B}_W(X,0)e^{i\overleftarrow{H}^m_{\Lambda}t}\Big)
\end{equation} where $\overrightarrow{H}^m_{\Lambda} = \Big(1+\frac{\hbar\Lambda}{2i}\Big)\hat{H}_m$, and $\overleftarrow{H}^m_{\Lambda}$ is defined in a corresponding manner.

The next step is to decompose the forward and backward evolution operators in Eq. (15) into short-time segments, 
as is done in the path-integral representation of quantum mechanics\cite{fbts1}. Then, complete sets of coherent states are used 
to expand the forward and backward evolution operators in each short-time interval. These sets of coherent states, $|z\rangle$, act on the mapping space and can be defined through the following relations, $\hat{a}_{\lambda} | z \rangle = z_{\lambda} |z\rangle$, and
$\langle z| \hat{a}^{\dag}_{\lambda} = z^{*}_{\lambda}|z\rangle$, where $|z\rangle = |z_1,..., z_N\rangle$, 
with eigenvalue $z_{\lambda} = ({q}_{\lambda} + ip_{\lambda})/\sqrt{2}$. The continuous variables $q = (q_1, ..., q_N)$ and $p = (p_1, ..., p_N)$ are the mean coordinates and momenta of the harmonic oscillators in the coherent state $|z\rangle$, respectively. 

If one assumes that coherent state variables at subsequent time-slices are orthogonal, i.e. $\langle z_j(t)|z_{j+1}\rangle \approx \pi^N \delta (z_{j+1} - z_j(t_j))$, the time evolution that is generated by the mapping Hamiltonian on the coherent states can be represented by a continuous trajectory evolution in the $(q, p)$ phase space\cite{fbts1}. This approximation is the essential step in deriving the FBTS evolution equations below, and can be systematically relaxed by introducing Monte Carlo sampling of new coherent state variables at each new time slice\cite{fbts2}, however this procedure can be numerically costly. Upon making the orthogonality approximation, one performs a final scaling of the coherent state variables by a factor of $\frac{1}{\sqrt{2}}$, such that the continuous trajectories are generated by Hamiltonian equations of motion, that define the FBTS solution to the QCLE.
\begin{eqnarray}
\label{fbts-eq}
\frac{dq_{\mu}}{dt} = \frac{\partial H_{e}(X,x,x')}{\partial p_{\mu}} , \quad \frac{dp_{\mu}}{dt} = -\frac{\partial H_{m}(X,x,x')}{\partial q_{\mu}}, ~~~~~~ \nonumber \\
\frac{dq'_{\mu}}{dt} = \frac{\partial H_{m}(X,x,x')}{\partial p'_{\mu}} , \quad \frac{dp'_{\mu}}{dt} = -\frac{\partial H_{m}(X,x,x')}{\partial q'_{\mu}}, ~~~~~~ \\
\frac{d R}{dt} = \frac{P}{M}, \quad \frac{d P}{dt} = -\frac{\partial H_{m}(X,x,x')}{\partial R}, \hspace{2cm}\nonumber
\end{eqnarray}
where $z_{\mu} = (q_{\mu}+ip_{\mu})/\hbar$ are the eigenvalues of the $2N$ scaled coherent states that represent the subsystem. 
The Hamiltonian function that generates the FBTS evolution is
\begin{eqnarray}\nonumber
H_e(X,x,x') = H_{vib}(R,P) + \frac{1}{2\hbar}\sum_{\lambda \lambda'} h_{\lambda \lambda'}(R)\\ \times (q_{\lambda}q_{\lambda'}+p_{\lambda}p_{\lambda'}+q_{\lambda}'q_{\lambda'}'+p_{\lambda}'p_{\lambda'}'),	    
\end{eqnarray} where $(X,x,x') = (R,P,q,q',p,p')$. 

The formal expression for the average value of a (partially Wigner transformed) time-dependent operator can be written as follows, 
\begin{equation}
    \langle B (t) \rangle = \sum_{\lambda \lambda'}\int dX B^{\lambda \lambda'}_W(X,t) \rho^{\lambda' \lambda}_W(X). 
\end{equation} 

In the FBTS simulation algorithm, the matrix elements of the operator $\hat{B}_W(t)$ in the integrand above are approximated using the following expression,
\begin{eqnarray}\nonumber 
B^{\lambda \lambda'}_W(X,t) &=& \sum_{\mu \mu} \int dx dx' \phi(x) \phi(x') (q_{\lambda} + i p_{\lambda})(q_{\lambda}' - i p_{\lambda}') \\&&\nonumber \times B^{\mu \mu'}_W(X_t) (q_{\mu}(t) + i p_{\mu}(t))(q_{\mu'}'(t) - i p_{\mu'}'(t)),\\
\end{eqnarray}
where $\phi(x) = \hbar^{-N} e^{-\sum_{mu}(r_{\mu}^2+p_{\mu}^2)/\hbar}$ are normalised Gaussian distribution functions. 

Evaluation of the integrals over the time-independent $\phi$ functions in Eq. (19) is carried out by Monte Carlo sampling. 
Thus, implementation of the FBTS methods requires sampling of initial conditions from $\hat{\rho}_W(0)$, and the propagation of classical-like trajectories in the extended phase space the system. 

\subsection{FBTS Dynamics Simulations}
The initial state of the total (electron-vibrational) density operator is separable; a single electronic excitation is present on a donor molecule at the interface, and each set of vibrational DOF's on the lattice is in a canonical equilibrium state with temperature $T=300K$, such that, 
\begin{equation}
\hat{\rho}(0) = \hat{\rho}_{vib}^{eq} \otimes \hat{P}_{XT},  
\end{equation} where $\hat{P}_{XT}$ is a projector onto the interfacial exciton state. 

To initiate each trajectory, the forward and backward subsystem coordinates, $x$ and $x'$, are independently sampled from the $\phi$ distributions, and the bath initial conditions, $X_0=(R_0,P_0,)$, are sampled from the Wigner transform of the initial canonical density operator,
\begin{eqnarray}\nonumber
    &\rho_{vib,W}^{eq}(R,P) = \prod_{m,k}\frac{\tanh(\beta \omega_{m,k}/2)}{\pi} \\
    &\ \times \exp\Bigg[ - \frac{\tanh(\beta \omega_{m,k}/2)}{\omega_{m,k}}\Big[ P_{m,k}^2 + \omega_{m,k}^2 R_{m,k}^2\Big]\Bigg],
\end{eqnarray} with inverse temperature $\beta = (k_BT)^{-1}$.  

The real-time dynamics of exciton dissociation, charge transport, and charge separation can be monitored via time-dependence of operators such as individual site populations or the average distance between the electron and hole on the lattice. For the 0D model, for example, the latter of these quantities is
\begin{equation}
\langle L (t) \rangle  = \sum_{n=1}^N n L \rho_{nn}(t),
\end{equation} 
where the lattice spacing $L = 1.2$ $nm$, and $\rho_{nn}(t)$ are the diagonal elements of the time-dependent electronic reduced density matrix, and site 0 corresponds to the XT site, which does not contribute to the electron-hole separation. 

\section{Simulation Results}

\begin{figure}
  \includegraphics[width=\columnwidth]{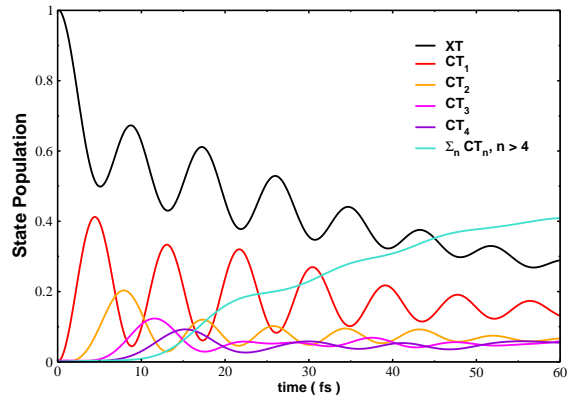}
  \caption{Time evolution of the XT and CT state populations in the 0D model with an effective Coulomb binding energy of $E_{Coul}^{eff} = \frac{e^2}{4\pi \epsilon_r\epsilon_0 L} = 0.30 eV$, and electronic couplings of $\Delta_{XT-CT}=0.15eV$, and $\Delta_{CT-CT}=0.10eV$. The XT site energy, is set to be nearly resonant with the delocalized CT states, $\varepsilon_{XT}=0$. The polaron formation time for the vibrational bath is set to $\tau = 20fs$, and the site reorganization energies are $\lambda=0.02 eV$, and $T=300K$.}
\label{1}
\end{figure}

First, we report results of FBTS dynamics simulations of the XT dissociation process in the zero-dimensional model, with zero applied field, in a parameter regime that is representative of functional OPV materials at room temperature\cite{tamura2013jacs,troisi2015,beljonne16,hodgkiss2014,kato2018}. In the zero-dimensional case, a single interfacial donor site (see Fig. 1) is coupled to a chain of 10 acceptor sites. Each lattice site is coupled to 200 vibrational modes, although the results reported here were found to converge with about 40 vibrational modes per site. Convergence with respect to the number of lattice sites was checked with up to $N=20$ for the 0D model. The simulation results for the zero dimensional model depicted in Fig. 2 correspond to ensemble averages containing $N_{traj} = 1x10^6$ FBTS trajectories. This ensures that the statistical error in the average state populations is rather small in this case. However, on longer timescales, and in systems with higher dimensionality, this level of convergence is not necessary to gain insight into the performance of the FBTS approach. Hence, slightly smaller statistical ensembles were used in most other simulations. Hence, where appropriate, the approximate size of the uncertainty is indicated by the size of representative points on the FBTS curves. 

\begin{figure}
  \includegraphics[width=\columnwidth]{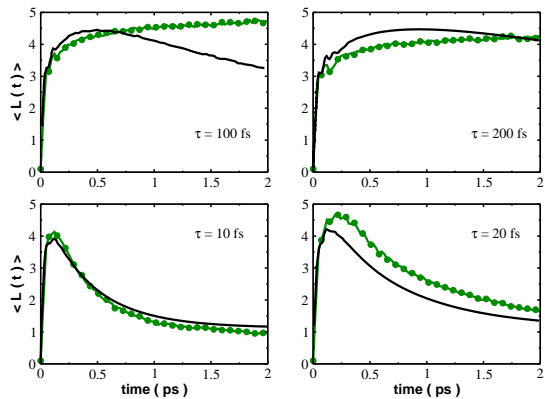}
  \caption{Time evolution of the mean electron-hole distance, $\langle L (t) \rangle$, in the zero-dimensional model, at zero applied field ($E = 0$). The value of the polaron formation time, $\tau = 1/\omega_c$,is labelled in each panel, and all parameters are the same as in Fig. 1. The black lines correspond to HEOM results from reference [34]. FBTS simulation results are shown in green; the size of the green dots on the FBTS curves represents the approximate size of the statistical error with $N_{traj}=5x10^4$.}
\label{2}
\end{figure}

In Fig. 2 the short-time dynamics of exciton dissociation are depicted. At $t=0$ the XT state is fully populated. As time proceeds the XT state coherently transfers probability to the nearest CT state, $CT_1$. This process initiates a cascade of population transfer along the lattice of CT sites. Charge separated (CS) states are defined here as being CT states that are separated by 4 or more lattice sites. Figure 2  shows that the FBTS simulations predict that the total population of the CS states rises to a substantial fraction on a $50$ $fs$ timescale. 

In order to gain further insight into both the initial fast charge delocalization, and the subsequent slower polaron formation process, we investigate the average spatial separation of the electron and hole on the timescale of a few picoseconds. In this case, direct comparisons can be made with the recent work of Kato and Ishizaki\cite{kato2018}, who have applied the highly accurate HEOM technique to this model. In Fig. 3, the mean electron-hole distance is plotted as a function of time, for a selection of different polaron formation times. The FBTS results nicely capture the fact that the short time-scale spatial delocalzation is independent of $\tau$ in this parameter range. For intermediate polaron relaxation times the FBTS results retain most of the qualitative features of the evolution, particularly at short times, but fall out of quantitative agreement with the benchmark HEOM results. 

\begin{figure}
  \includegraphics[width=\columnwidth]{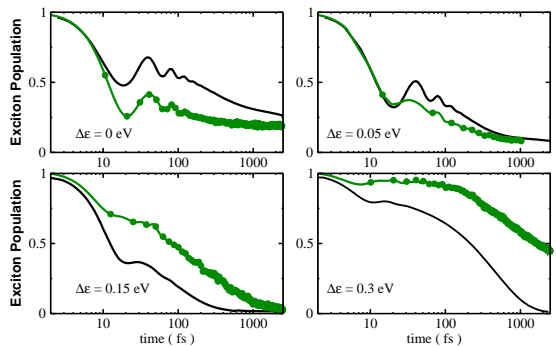}
  \caption{Time evolution of the interfacial exciton state population, in the zero dimensional model with a static electric field of $E = 10 V/\mu m $, for different values of the energetic bias, $\Delta \varepsilon = E_{XT} - E_{CT_1}$. In this case the effective Coulomb binding energy is $E_{Coul}^{eff} = 0.16 eV$, and the electronic couplings are $\Delta_{XT-CT}=0.05eV$, and $\Delta_{CT-CT}=0.05eV$. The polaron formation time for the vibrational bath is $\tau = 130fs$, and the site-reorganization energies are all set to $\lambda=0.05 eV$, with $T=300K$. The HEOM results from [50] are depicted in black, and FBTS simulation results are in green. The size of the green dots on the FBTS curves represent the size of the statistical error with $N_{traj}=5x10^4$. }
\label{3}
\end{figure}

Next, the effect varying the energetic bias is investigated in the presence of an applied field. Recently, Yan, Song, and Shi, have reported HEOM results for a version of the 0D model with parameters chosen to reproduce their quantum free energy calculations on this system\cite{shi2108}. In Figs. 4 - 6, we display how well the FBTS results can capture the effect of the energy bias on the charge transfer process, in the presence of a static electric field. This case provides a rather stringent test for this approximate dynamics approach as there are a number of competing energy scales, length scales, and time-scales.   

\begin{figure}
  \includegraphics[width=\columnwidth]{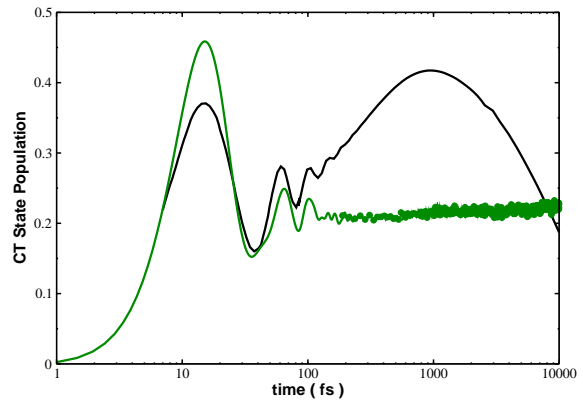}
  \caption{Time evolution of the interfacial CT state population, in the zero dimensional model with a static electric field of $E = 10 V/\mu m $, at zero energetic bias, $\Delta \varepsilon = E_{XT} - E_{CT_1} = 0$, and the other parameters are the same as Fig. 4. The HEOM results from reference [50] are depicted in black, and the FBTS simulation results are shown in green. The size of the green dots on the FBTS curves represent the approximate size of the statistical error with $N_{traj}=5x10^4$. }
\label{4}
\end{figure}

In Fig. 4, FBTS time evolution of the XT state is plotted at zero bias, and for three different values of excess XT energy. Benchmark data from the HEOM method of Ref. [50] are again shown for comparison. In all cases shown in Fig. 4 the FBTS simulations are accurate at short times, and qualitatively capture most aspects of the XT decay process out to timescales of about a few picoseconds. Figure 5 shows the time evolution of the interfacial CT state population at zero bias. Again, while FBTS performs well on the sub-picosecond timescale, it misses the revival of the CT state that occurs before the charge separation is complete. Inaccuracies in state-populations of this size and are not uncommon in nonadiabatic semiclassical and quantum-classical methods\cite{mfgqme}, and usually indicate some violation of the detailed balance condition has occurred in the dynamics. 

Next, the timescale of charge separation is investigated. Figure 6 shows the time evolution of the total population of the CS states in the zero dimensional model, with an XT energy bias of $0.30$ $eV$. The corresponding XT state evolution is plotted in the bottom-right panel of Fig. 3. The onset of charge separation in the FBTS results is delayed due to the incomplete XT dissociation at this value of the bias seen in Fig.3. However, once the CS process initiates, the FBTS simulations give a reasonable representation of the CS process. 

\begin{figure}
  \includegraphics[width=\columnwidth]{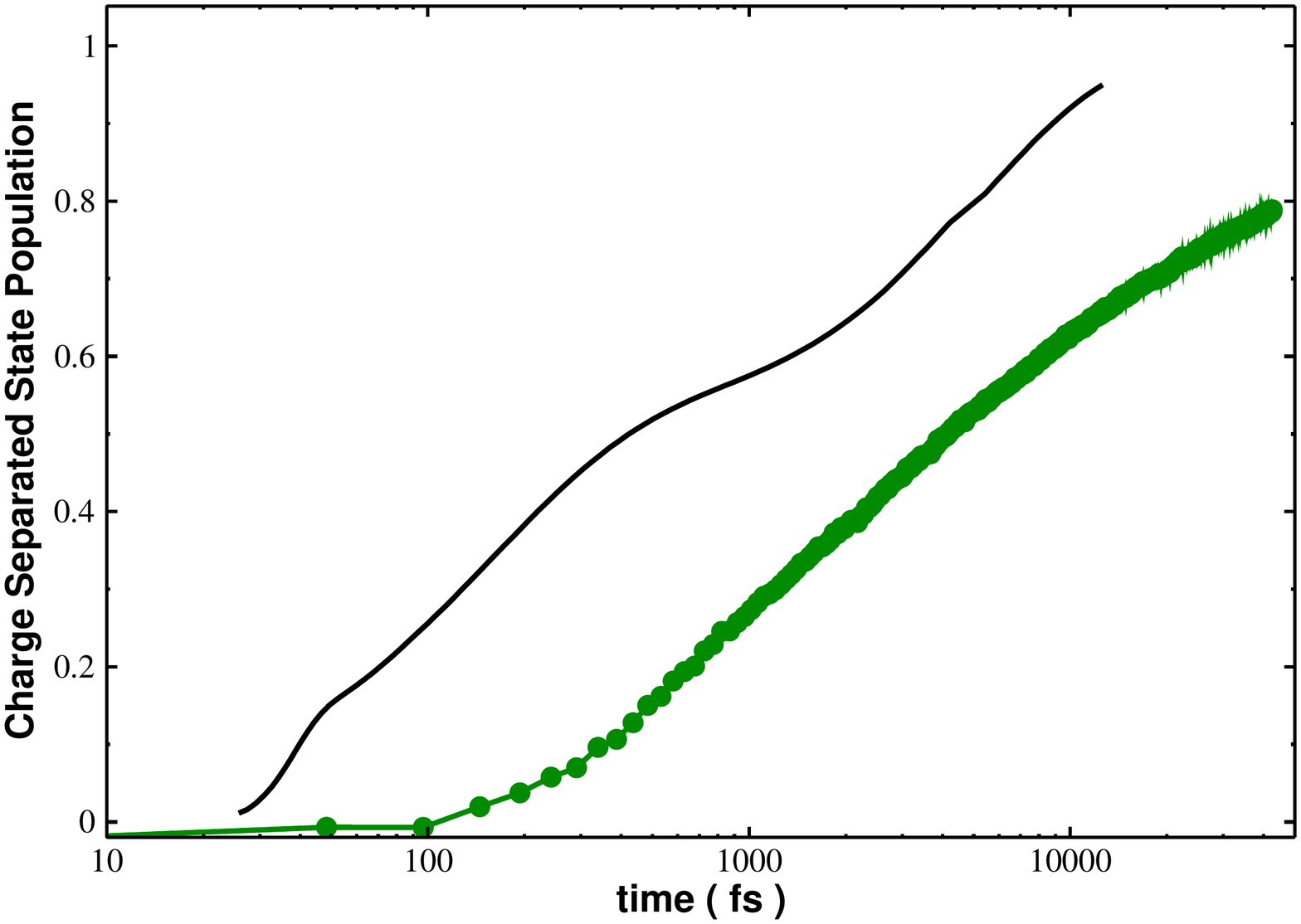}
  \caption{Time evolution of the charge separated state population, in the zero dimensional model with a static electric field of $E = 10 V/\mu m $, at an energetic bias of $\Delta \varepsilon = E_{XT} - E_{CT_1} = 0.30 eV$. The other parameters are the same as Fig. 4. The HEOM results from reference [50] are depicted in black, and FBTS simulation results are shown in green. The size of the green dots on the FBTS curves represent the approximate size of the statistical error with $N_{traj}=5x10^4$.}
\label{6}
\end{figure}

We now move on to briefly investigate the XT dissociation dynamics in higher dimensional cases. Such cases are much less accessible to numerically accurate quantum dynamics solvers due to the fact the size of the CT Hilbert space grows like $N_s^2$, where $N_s$ is the number of donor or acceptor sites on the lattice, while the number of vibrational degrees of freedom in the problem scales as $N_{vib} N_s^2$. Hence the size of the problem can grow extremely rapidly, and calculations can easily become prohibitively time consuming; even for simple trajectory-based quantum classical dynamics techniques. In Fig. 7 the average electron-hole distance, $\langle L (t) \rangle$ is plotted for the one-dimensional model, with zero energetic bias, and at zero applied field ($E = 0$). The total number of donor and acceptor sites on each side of the interface is $N=8$, and there is a single exciton site at the interface. All other parameters are chosen to be the same as in Fig. 4. Similar to the case of the 0D model, fast delocalization across the CT manifold is observed, with an average charge separation of approximately 4 or 5 lattice sites being established on a picosecond timescale.  

\begin{figure}
  \includegraphics[width=\columnwidth]{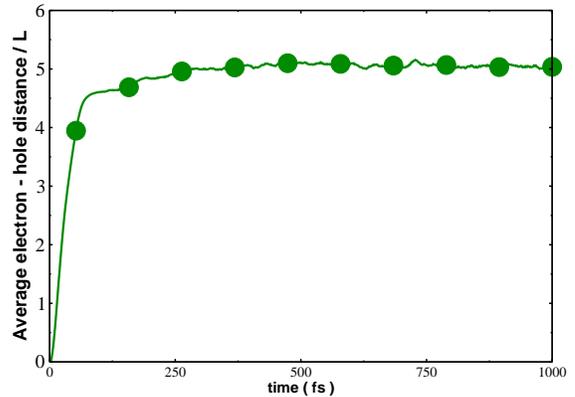}
  \caption{Time evolution of the mean electron-hole distance, $\langle L (t) \rangle$, in the one-dimensional model, with zero energetic bias, and at zero applied field ($E = 0$). All other parameters are the same as in Fig.4. The size of the green dots on the FBTS curve represents the size of the statistical error with $N_{traj}=1x10^5$.}
\label{7}
\end{figure}

Finally, in Fig. 8 the short time evolution of the XT and CT populations for the 1D model and the 2D model are shown. A large fraction of the XT population transfers through the CT manifold on this ultrafast timescale. As evidenced in the average electron-hole pair separation, this population is transferred not only to the nearest interfacial CT state, but to further delocalized CT states as well. This indicates that the fast delocalization mechanism, even at zero energetic bias, is rather insensitive to the dimensionality of models studied here.  

\begin{figure}
  \includegraphics[width=\columnwidth]{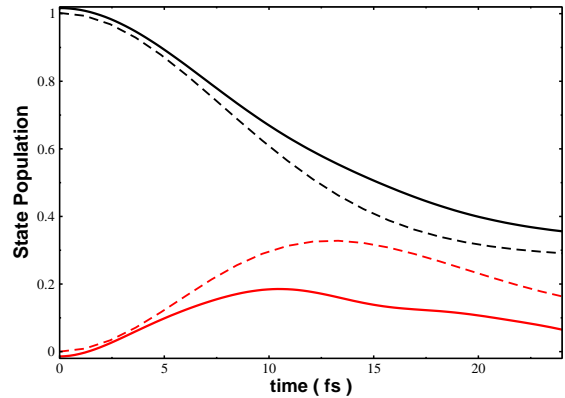}
  \caption{Time evolution of the initial XT (black lines) and nearest interfacial CT state (red lines) populations in the 1D (dashed lines) and the 2D (solid lines) models. The parameters are identical to Fig. 7. The 2D model comprised a single XT state in the middle of the interface, and grids with 36 total sites for each of the donor and the acceptor domains, 40 bath modes were included per site, and $N_{traj}=5x10^3$.}
\label{8}
\end{figure}

\section{Conclusions and Outlook}
This work is intended to pursue some of the initial, but essential, steps toward developing accurate and efficient quantum-classical dynamics simulation approaches that can be used to help understand nonequilibrium, time-dependent, charge and energy transport problems in condensed phase systems. Here, the focus was put on a class of tight-binding lattice models that are representative of the donor-acceptor interface of heterojunction OPV materials. In this setting, the process of XT dissociation, polaron formation, CT decay and long-range CS were investigated. In order to make contact with available highly accurate quantum mechanical data, many elements of the lattice models adopted here have been kept rather simple. However, owing to the flexibility of the quantum-classical dynamics approaches, there are no particular limitations in this regard. Hence, future work in this direction could include static disorder in the energy landscape, or non-local electronic couplings, for example. 

One of the major issues in moving toward simulating realistic, three dimensional, systems will be coping with the proliferation in the number of electronic and vibrational degrees of freedom. In this regard, new efficient simulation methods\cite{mfgqme,saller2019} still need to be developed, to help improve the accuracy and bring down the computational cost, to make long time-scale and large length-scale simulations more routinely tractable. 

In view of the performance of the FBTS simulations versus the benchmark HEOM data for the models studied here, the overall accuracy of this particular quantum-classical approach is rather encouraging. FBTS offers sufficient accuracy to obtain a number of cursory insights into the different time-scales involved in the passage of probablility through the manifold of XT, CT, and CS states in these models. It also nicely reproduces some of the quantitative features of the XT delocalization and CS process. However, the FBTS method alone is likely incapable of resolving fine details of the transport mechanism that occur on time-scales longer than a few picoseconds, ultimately owing to an incorrect description of detailed balance in the dynamics. In principle, however,the accuracy of the FBTS approach can be improved in combination with the generalized quantum master equation\cite{win}, by introducing stochastic sampling of "jumps" in the mapping degrees of freedom along the trajectories\cite{fbts2}, or perhaps via other means\cite{saller2019}. Hence, these initial FBTS results indicate a number of routes toward refining the tool-box of quantum-classical dynamics approaches for further use in simulating charge and energy transfer in condensed phase systems.       

\section*{Acknowledgements}
AK acknowledges start-up funds from Dalhousie University, and support from the NSERC Discovery Grant Program. 

\bibliography{arXiv_version}

\end{document}